\documentclass[aps,pre, twocolumn, showpacs, superscriptaddress, floatfix]{revtex4}

\usepackage{amssymb, amsmath, amsfonts}
\usepackage{bm}
\usepackage{graphicx}

\usepackage{subfigure}

\begin{document}

\title{Role of the particle's stepping cycle in an asymmetric exclusion process: A model of mRNA translation} 

\date{14 December 2009}

\author{L. Ciandrini}
\email{l.ciandrini@abdn.ac.uk}
\affiliation{Institute for Complex Systems and Mathematical Biology, King's College,
University of Aberdeen, AB24 3UE Aberdeen, United Kingdom}
\author{I. Stansfield}
\affiliation{Institute of Medical Sciences, Foresterhill, University of Aberdeen, AB25 2ZD Aberdeen, United Kingdom}
\author{M. C. Romano}
\affiliation{Institute for Complex Systems and Mathematical Biology, King's College,
University of Aberdeen, AB24 3UE Aberdeen, United Kingdom}
\affiliation{Institute of Medical Sciences, Foresterhill, University of Aberdeen, AB25 2ZD Aberdeen, United Kingdom}

\begin{abstract}
Messenger RNA translation is often studied by means of statistical-mechanical models based on the Asymmetric Simple Exclusion Process (ASEP), which considers hopping particles (the ribosomes) on a lattice (the polynucleotide chain). 
In this work we extend this class of models and consider the two fundamental steps of the ribosome's biochemical cycle following a coarse-grained perspective. In order to achieve a better understanding of the underlying biological processes and compare the theoretical predictions with experimental results, we provide a description lying between the minimal ASEP-like models and the more detailed models, which are analytically hard to treat.
We use a mean-field approach to study the dynamics of particles associated with an internal stepping cycle.
In this framework it is possible to characterize analytically different phases of the system (high density, low density or maximal current phase). 
Crucially, we show that the transitions between these different phases occur at different parameter values than the equivalent transitions in a standard ASEP, indicating the importance of including the two fundamental steps of the ribosome's biochemical cycle into the model.
\end{abstract}

\maketitle

\section{Introduction}

The translation of the messenger RNA (mRNA) is the final step of protein synthesis. During this process the information enclosed in the triplet code of the nucleotide chain is \emph{translated} into the amino acid sequence of the encoded proteins. Translation is usually viewed as a three-stage process~\cite{Alberts:2008qp, Lodish:yq, Kapp:2004ty}: during \emph{initiation} a ribosome (complex of proteins and RNA) binds the mRNA molecule (a sequence of nucleotides previously transcribed from the DNA) and after a series of biochemical reactions, it moves along the chain. This stage in which the protein is built up amino acid by amino acid according to the mRNA sequence is called \emph{elongation}. Each elongation step consists in turn of a series of biochemical reactions which define the ribosome's biochemical cycle. Lastly, the ribosome reaches the termination codon and leaves the mRNA releasing the protein. This last step is called \emph{termination}. In this paper we propose a model for the elongation stage of mRNA translation. 

This process, primarily controlled by the dynamics of ribosomes along the mRNA chain, bears a resemblance to a one-dimensional driven lattice gas. For this reason, the mRNA translation inspired a statistical-mechanical class of models known as Asymmetric Simple Exclusion Processes (ASEPs). They have been introduced in the biophysical literature as models representing the dynamics of ribosomes along an mRNA chain \cite{MacDonald:1968tg, MacDonald:1969rt}. Later, this class of model has been studied from a more theoretical point of view \cite{Derrida:1992by, Derrida:1993it, Derrida:1998la, Schmittman:1995dn, Schutz:2001ms, Parmeggiani:am, Lakatos:2003rt, Harris:2004mq, Shaw:2004ty, Dong:2007dq, Blythe:2007tw} and the possible biological applications have been rediscovered only recently, not only for protein synthesis \cite{Lakatos:2003rt, Shaw:2003xw, Chou:2004bh, Dong:2007fc, Romano:2009hc} but also for the movement of molecular motors \cite{Chowdhury:2005rr, Pierobon:2006mz, Pierobon:2009vo}. Other non-biological applications have been studied too (see, e.g.,~\cite{Chowdhury:2000rp}). A detailed discussion of the ASEP can be found, for instance, in Refs.~\cite{Schutz:2001ms, Blythe:2007tw}.

A traditional exclusion process consists of particles moving along a lattice with only steric interactions. In other words, each lattice site can be occupied just by one particle at a time. 
Although this approach is interesting from a theoretical point of view, it is however not a realistic way of describing translation since it encompasses the whole ribosomal elongation cycle in a single step. Other approaches to modelling translation include several (up to fifteen) distinct phases of the ribosome's mechano-chemical cycle~\cite{Zouridis:2007la, Basu:2007qw, Garai:2009ta}.
We want to place ourselves between the class of minimal models and the more detailed models, which are difficult to analyze. Consequently, we consider two fundamental steps of the ribosome's biochemical cycle following a coarse-grained picture (Section~\ref{sec::model}). We show that former ASEP-like models correspond to a limiting case of the model we use. As will become clear from the discussion, this limiting case is however biologically not plausible, showing the need for an extension of previous models.

In the following section we explain the biological framework (in particular the role of ribosomes and transfer RNAs) and then in Section~\ref{sec::model} we introduce the model from a mathematical point of view. The results for periodic and open-boundary systems are presented in Section~\ref{sec::results}. Using the same approach as Ref.~\cite{Derrida:1992by} we analyze the model in the mean-field approximation and compare this model with a typical exclusion process. In addition, we show that the same results can be achieved by using an extremal principle~\cite{Krug:1991tg, Popkov:1999hc, Hager:2001hw}. Although we observe the same variety of phase transitions that one would expect from the simpler case without the particle's internal states, the locations of the critical points change substantially and depend on the internal dynamics of the particles. Finally, in Section~\ref{sec::conclusions} we discuss the results and the conclusions from a theoretical and biological viewpoint.

\section{Biological background\label{sec::biology}}
Here we briefly introduce the underlying biological process that we want to describe. More information can be found, e.g., in Refs.~\cite{Alberts:2008qp, Lodish:yq, Kapp:2004ty}.

\begin{figure}[htbp]
\begin{center}
\includegraphics[width=0.45\textwidth]{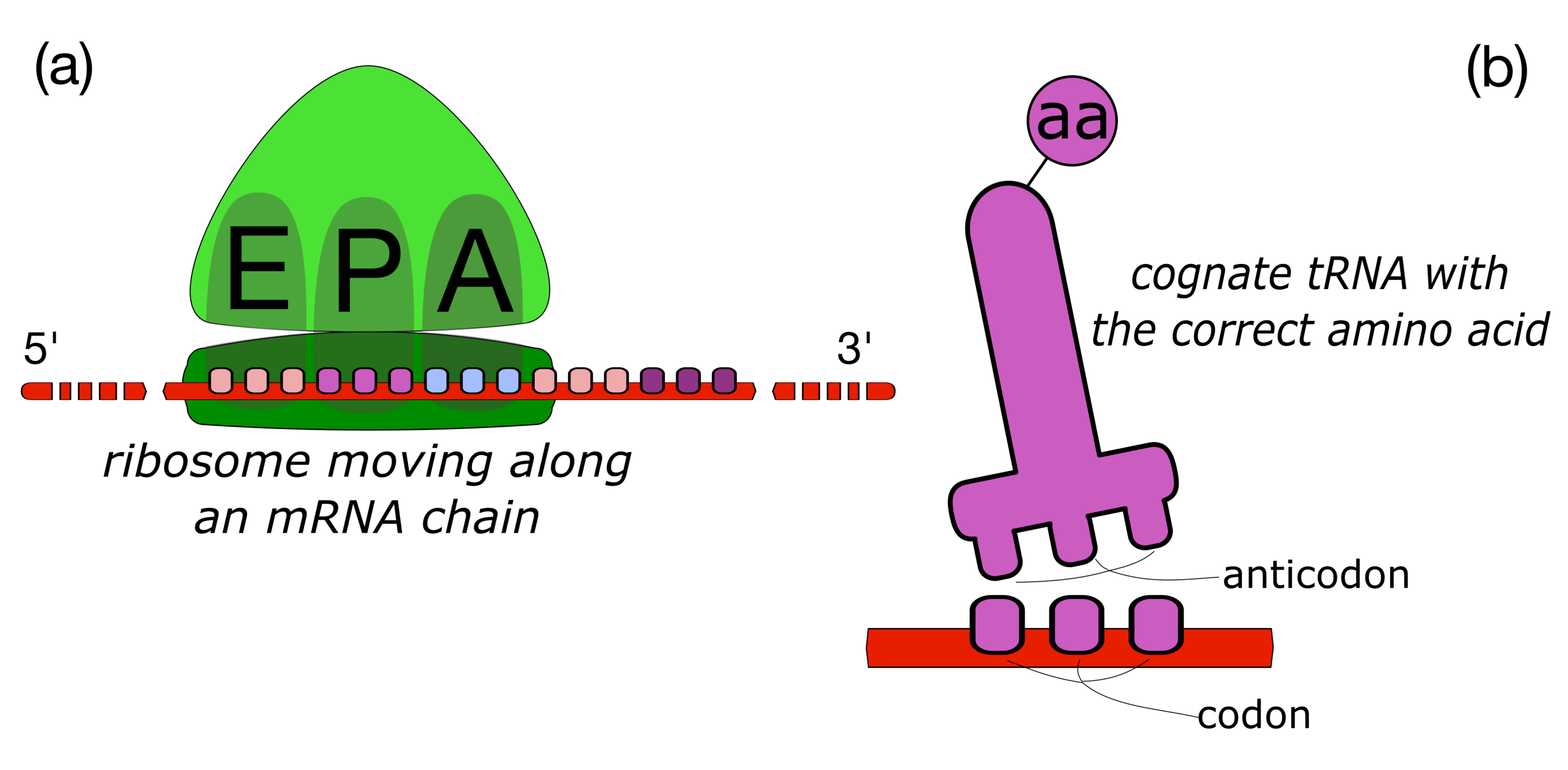}
\caption{(a) Illustration of a ribosome with empty A, P, and E sites along an mRNA chain. (b) Classical representation of a tRNA with its anticodon in the lower region and the corresponding amino acid (aa) bound together.}
\label{fig::1}
\end{center}
\end{figure}

The mRNA is a nucleotide chain composed of four different bases (A,C,G,U); each group of three nucleotides is called a codon and specifies a certain amino acid.  
The keys for deciphering the code (the sequence of codons) are the transfer RNAs (tRNAs), freely diffusing molecules carrying amino acids. They have a region composed of three nucleotides (the anticodon) matching the corresponding codon on the mRNA. Moreover, tRNAs with the same anticodon transport the same amino acid. In most species, there are 35-40 distinct species of tRNA, each responsible for decoding a particular set of codons.
Ribosomes are complexes of proteins and RNA which move along the mRNA in a fixed direction (from the 5' to the 3' region, i.e. from the part of the chain that has been transcribed first towards the other end) and catalyze the assembly of amino acids delivered by tRNAs. Each ribosome has three regions of interaction for the tRNA. They are called Aminoacyl (A), Peptidyl (P)  and Exit (E) sites (see Fig.~\ref{fig::1}).

\begin{figure}[htbp]
\begin{center}
\includegraphics[width=0.45\textwidth]{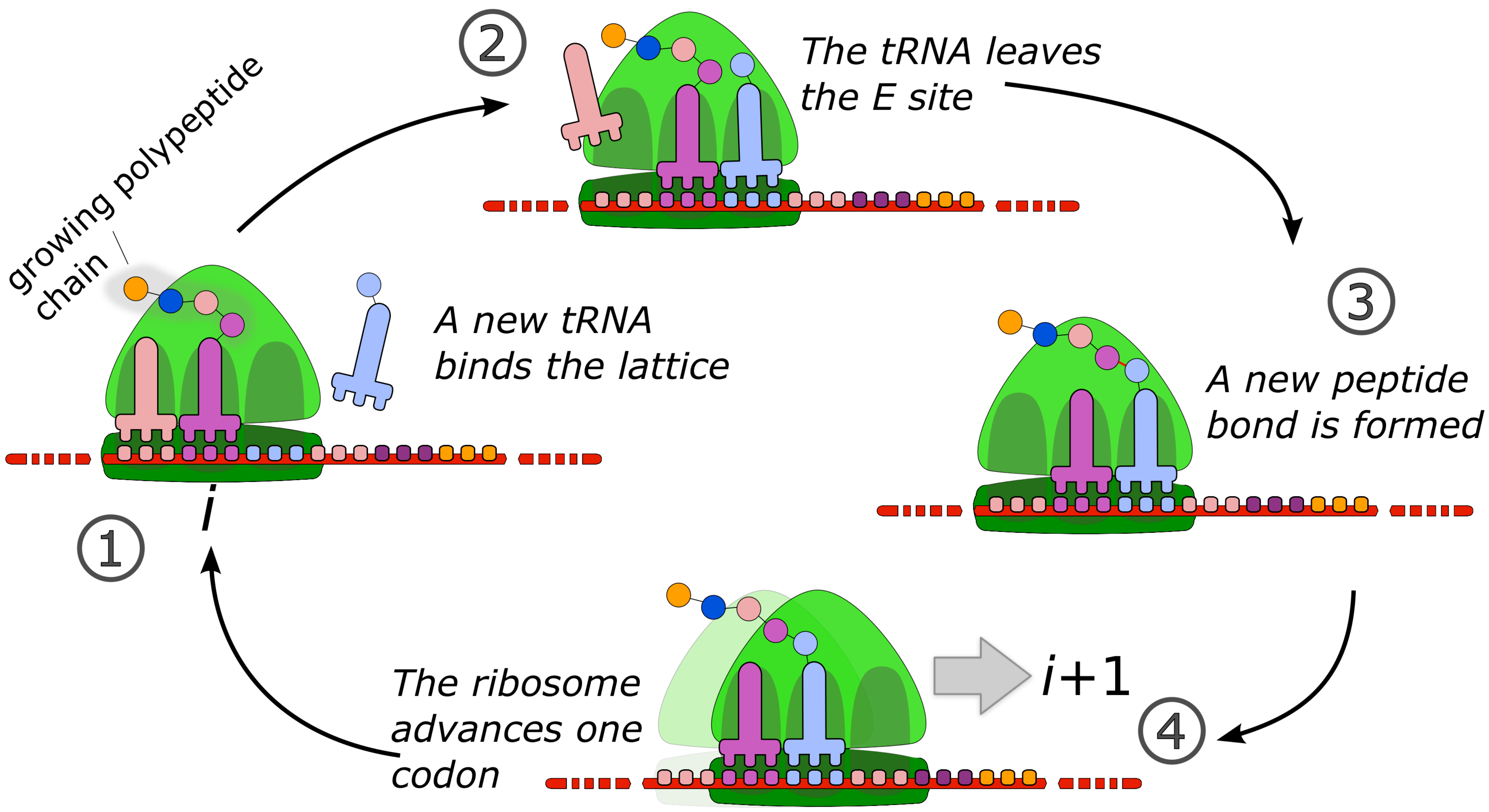}
\caption{Sketch of the translation elongation process. Once the ribosome finds the cognate tRNA (1), the tRNA in the E site abandons the ribosome (2) and the peptide carried by the existing, P-site tRNA binds the amino acid on the new, A-site tRNA (3). At that point the ribosome translocates provided that the next codon is empty (4). The ribosome is thus in the position to accept another tRNA and iterate the elongation till the end of the mRNA chain.}
\label{fig::2}
\end{center}
\end{figure}

The main steps of the elongation process are shown in Fig.~\ref{fig::2}. Following the translation initiation, a tRNA  is in the P site bound with the first amino acid of the growing polypeptide chain. Then, a complex of EF1$\alpha$ $\cdot$ GTP $\cdot$ tRNA$^{aa}$ (with tRNA$^{aa}$ we denote a tRNA bound to the amino acid $aa$) diffuses into the empty A site on the codon at position $i$. If the anticodon of the tRNA cannot base-pair with the codon, then the complex EF1 $\cdot$ GTP $\cdot$ tRNAaa is released and the process is repeated until a correct tRNA binds the ribosome. If on the other hand the correct tRNA anticodon base-pairs with the corresponding codon on the mRNA, then GTP is hydrolyzed and a conformational change in the ribosomal structure occurs. This change leads to the transfer of the nascent peptide from the P-site tRNA, to the amino acid carried by the A-site tRNA. The altered structure of the ribosome does not allow the cognate tRNA to unbind and leave the chain. Following the peptidyl transfer reaction and the incorporation of the new amino acid into the growing polypeptide chain, the ribosome translocates one codon (assuming ribosome progress is not blocked by any stalled ribosomes at downstream positions on the mRNA). The translocation process is catalyzed by the complex EF2 $\cdot$ GTP which induces a second conformational change in the ribosome. The tRNA at the P site is then transferred to the E site. The ribosome is now back to the first step of its biochemical cycle, with the growing polypeptide chain bound to the tRNA in the P site and the empty A site on the codon $i+1$, ready to receive another tRNA complex.
This process is iterated until the end of the mRNA chain, where the ribosome disassociates from the system and releases the synthesized protein. As soon as initiating ribosomes have moved sufficiently downstream to create space at the beginning of the mRNA, a new ribosome can bind the polynucleotide chain. Thus, several ribosomes can translate the same mRNA at the same time. 

Experimental data strongly indicate that searching for the correct tRNA, and not the translocation, is the rate limiting step of the biochemical cycle of the ribosome~\cite{Bilgin:1988tg, Schilling-Bartetzko:1992fp}. Therefore, in this work we shall approximate the whole biochemical cycle of the ribosome by a two-state cycle: (i) \emph{searching} for the correct tRNA and (ii) \emph{translocation} from one codon to the next. 
 
\begin{figure}[htbp]
\begin{center}
\includegraphics[width=0.45\textwidth]{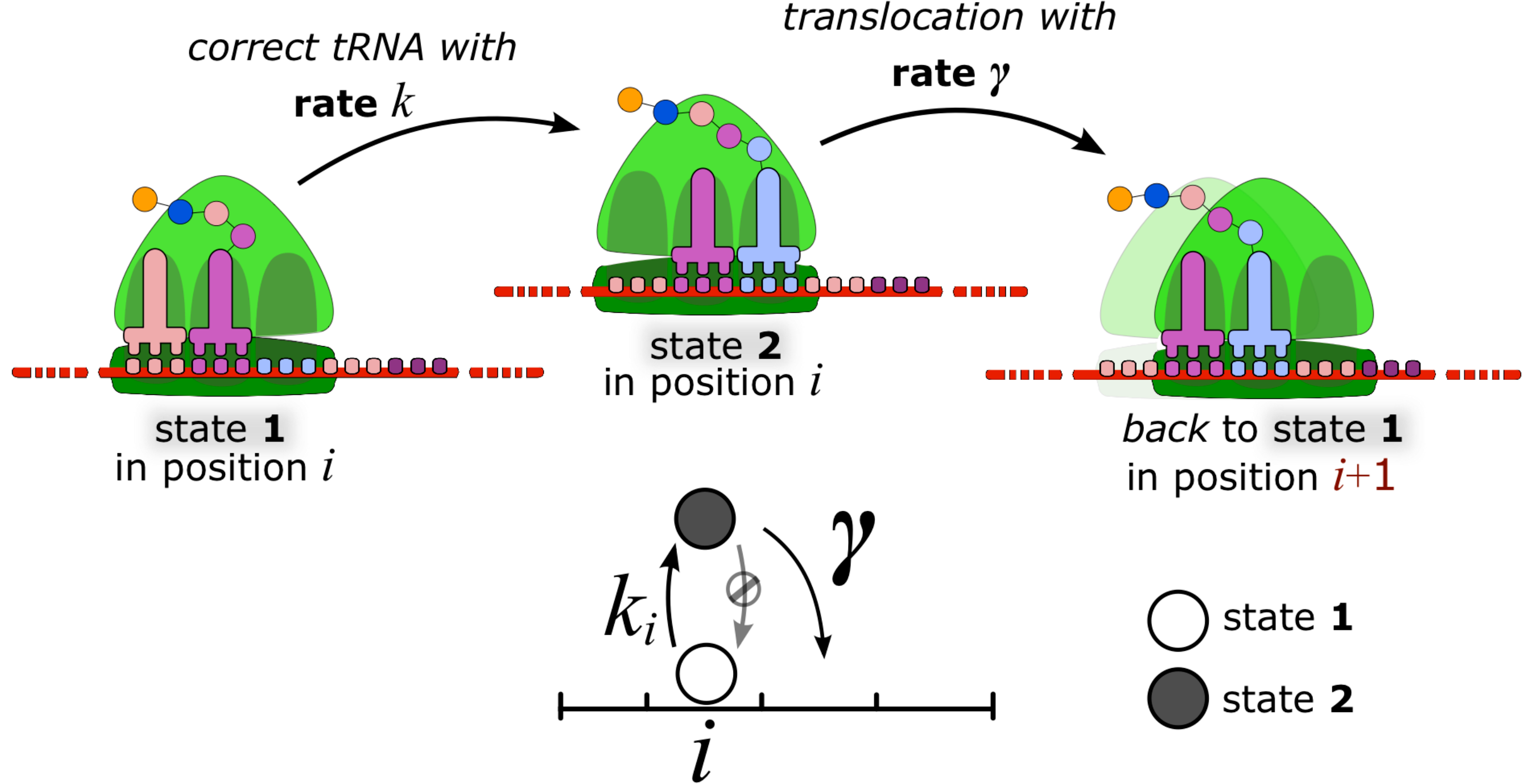}
\caption{The two states that a particle assumes represent a ribosome waiting for the cognate tRNA (state 1) and ready to translocate (state 2). The transition from ``state 1'' to ``state 2'' occurs with rate $k$ (in general depending on the codon $i$) which mainly models the concentration of tRNA$^{aa}$.  The translocation occurs with rate $\gamma$. The transitions are not reversible.}
\label{fig::3}
\end{center}
\end{figure}

From the modelling point of view, we shall consider particles changing an internal degree of freedom, or state, which influences their motion. Thus, a ribosome is represented by a two-state particle denoting the absence of the cognate tRNA in its Aminoacyl site (state $1$) or its presence (state $2$) as outlined in Fig.~\ref{fig::3}.

Throughout this work we assume that the concentration of charged tRNAs is homogeneous and large enough to neglect fluctuations (the transition rates do not change with time). The effects of limited resources~\cite{Cook:2009ez} and their 3D diffusion in the cytoplasm~\cite{Parmeggiani:2009yj} are not taken into consideration here. The individual charged tRNA concentrations govern the transition rates, which in general depend on the particular type of codon.

\section{The model \label{sec::model}}
We describe the mRNA molecule as a lattice of discrete sites, each one representing one codon. Ribosomes are represented by particles hopping from one site of the lattice to the next. 
From a more mathematical perspective, the occupation number $n_i=0,1,2$ of the site $i$ describes the different states in which it can be found. We say that a site is empty if its occupation number is $0$. A given site $i$ occupied by a particle in state $1$ or $2$ is respectively described by $n_i =1$ or $n_i=2$. The set $\eta = \{n_1, \dots, n_L \}$, where $L$ is the length of the lattice, will give the configuration of the system. The only transitions allowed are the following:
\begin{subequations}
	\label{eq::rules}
	\begin{align}
	1 &\rightarrow 2   & \text{with rate $k_i$}\\
	20 &\rightarrow 01 & \text{with rate $\gamma$,}
	\end{align}
\end{subequations}
where the first line means that a particle in state $1$ at site $i$ changes into state $2$ with rate $k_i$, which in general depends on the site $i$. The second line is a schematic representation of the translocation of a particle in state $2$ to the next site. Notice that a hopping particle is carried back to the state $1$. The values of $n_i$'s change with time according to these dynamical rules.
This dynamics has been first introduced in the literature by Klumpp and coworkers in~\cite{Klumpp:2008cq} to model the traffic of molecular motors on a filament.

The mean density $\rho_i$ of particles on site $i$ can be written discerning the contribution of particles in state $1$ and particles in state $2$. Thus, we shall say that $\lambda_i$ is the mean density of particles in state $1$ at site $i$ and $\sigma_i$ is the analogous for particles in state $2$. One can write these densities in terms of the occupation numbers:
\begin{eqnarray*}
	\lambda_i &= &\left< n_i (2-n_i)\right> \\
	\sigma_i &= & \left< \frac{n_i (n_i-1)}{2}\right> \;.
\end{eqnarray*}
The mean density of particles at site $i$ is then given by $\rho_i = \lambda_i + \sigma_i$. The brackets indicate the average of the quantities over time.

The lattice in consideration may have periodic or open boundary conditions. We shall study these cases in Sections~\ref{ssec::periodic} and \ref{ssec::open}, respectively.

From now on we study the case in which $k_i = k$ $\forall i$ and use the mean-field approximation, i.e. we neglect correlations between the sites ($\langle n_i n_j \rangle \simeq \langle n_i \rangle \langle n_j \rangle$). 
These approximations simplify the analysis considerably and, as we shall show later, yield qualitatively the same results.

With these prescriptions, the mean-field equations describing the evolution of the densities at site $i$ read as follows:
\begin{subequations}
		\begin{align}
			\cfrac{d  \lambda_i}{d t}  & =  \sigma_{i-1}(1-  \lambda_i -  \sigma_i) \gamma - k \lambda_i \\
			\vspace{0.5ex}
			\cfrac{d  \sigma_i}{d t}   & =  k \lambda_i  - \sigma_{i}(1-  \lambda_{i+1} -  \sigma_{i+1}) \gamma \;.
		\end{align}
		\label{eq::densities}
\end{subequations}
The current $J_+^i$ ($J_-^i$) is defined as the number of particles entering (leaving) the site $i$ per unit time. We can write the expression of the incoming and outgoing currents by using Eqs.~(\ref{eq::densities}):
\begin{align*}
			J_+^i  & =  \sigma_{i-1}(1-  \lambda_i -  \sigma_i)\gamma \\
			J_-^i   & =   \sigma_{i}(1-  \lambda_{i+1} -  \sigma_{i+1})\gamma \;.
\end{align*}

In this work we consider the steady-state condition ($\frac{d  \lambda_i}{d t}  = \frac{d \sigma_i}{d t}   = 0 $ $\forall i$), where the currents are the same along the lattice ($J:= J_-^i = J_+^i$ $\forall i$).

\section{Results\label{sec::results}}
First we study the effects of two-state particles in closed lattices and then we investigate the boundary-induced phase transitions in open systems. The predictions of the mean-field theory are then compared to numerical simulations performed with a Bortz-Kalos-Lebowitz-like algorithm~\cite{Bortz:1975lq} modified for the dynamic rules (\ref{eq::rules}), i.e. a continuous-time Monte Carlo which uses a random sequential updating scheme.
The first $10^6$ iterations of the algorithm are disregarded. Then, with the system in the steady-state, data is collected every $100$ iterations, for a total number of $10^6$ iterations.

\subsection{Periodic-boundary conditions \label{ssec::periodic}}
Since all sites in a lattice with periodic-boundary conditions are identical (in the special case of $k_i=k$ $\forall i$), we write Eqs.~(\ref{eq::densities}) without the indices $i$. Note that the mean number of particles in state $1$ is equal to the local density $\lambda$, i.e., $L^{-1}\left<\sum_{i=1}^L n_i \delta_{n_i , 1}\right> = \lambda$. The same holds for particles in the upper state:  $(2L)^{-1} \left<\sum_{i=1}^L n_i \delta_{n_i , 2} \right>= \sigma$. One obtains the following equations for the current of particles $J$ and for the densities:
\begin{equation}
	\label{eq::J1}
	J  =  \sigma (1- \lambda -\sigma) \gamma \;,
\end{equation}  
\begin{subequations}
		\begin{align}
			\lambda  &=  \cfrac{J}{k}\\
			\sigma &= \rho - \lambda = \rho- \cfrac{J}{k}\;.
		\end{align}
		\label{eq::dens1}	
\end{subequations}

The density $\rho$ plays the role of the control parameter. For this reason, it is useful to write Eqs.~(\ref{eq::J1}) and~(\ref{eq::dens1}) as follows
\begin{equation}
	\label{eq::Jring}
	J  = \cfrac{ \rho (1-\rho) k}{\frac{k}{\gamma}+  (1-\rho)} \;,
\end{equation} 
\begin{equation}
	\begin{aligned}
		\lambda & =  \cfrac{ \rho (1-\rho)}{\frac{k}{\gamma} + (1-\rho)}\\
		\sigma & = \cfrac{ \frac{k}{\gamma} \rho}{\frac{k}{\gamma}+ (1-\rho)}\;.
	\end{aligned}
\end{equation}
 
\begin{figure}[hbtp]
	\centering
	 \includegraphics[width=0.28\textwidth]{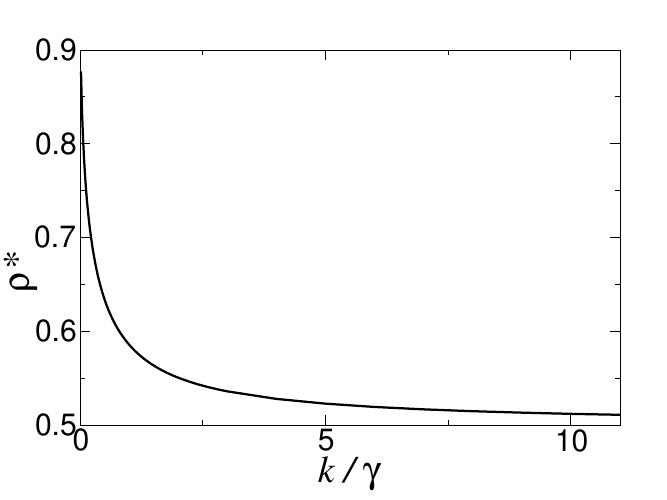}
	 \caption{Plot of $\rho^*$ as a function of $k/\gamma$, Eq.~(\ref{eq::rho_max}). \label{fig::rho_max}}
   \end{figure}

\noindent These results have been previously obtained in~\cite{Klumpp:2008cq}. Note that both densities $\lambda$ and $\sigma$ are functions of the ratio $k/\gamma$. This result is not an artifact of the mean-field approximation, since simulations confirm this dependence (data not shown). For the sake of simplicity, in simulations we can therefore set the value of $\gamma$ to a fixed value  (for instance $\gamma=1$).
  \begin{figure}[h!btp]
	\centering
	 \includegraphics[width=0.5\textwidth]{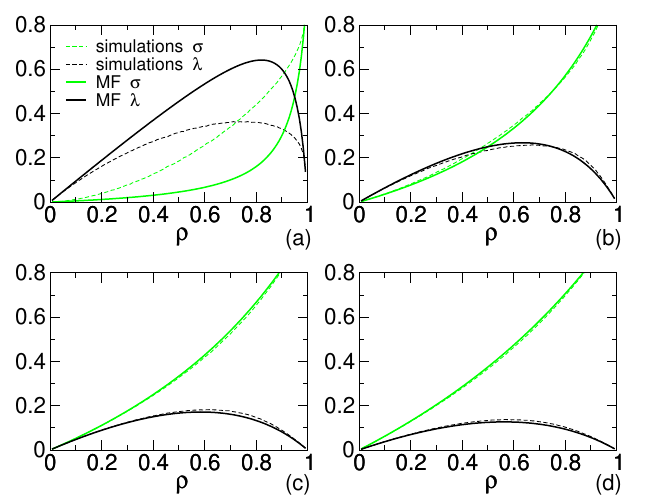}
	 \caption{(Color online). Simulations (dashed lines) and mean-field (MF) approximation (full lines) of the densities $\lambda$ in black and $\sigma$ in green (light gray). This figure shows the curves for a ring with $N=250$, $\gamma=1$ and $k=0.05$ (a), $k=0.5$ (b), $k=1$ (c), $k=1.5$ (d).  \label{fig::MF}}
   \end{figure}
From Eq.~(\ref{eq::Jring}) one can obtain the value of $\rho$ for which the current is maximal
\begin{equation}
	\rho^* :=1+  \frac{k}{\gamma} -  \frac{k}{\gamma} \sqrt{1+ \frac{\gamma}{k}} = 1- \chi \;,
	\label{eq::rho_max}
\end{equation}
where $\chi:=(k/\gamma)(\sqrt{1+\gamma/k}-1)$, and the maximal value of the density of particles in the state $1$ (proportional to the current $J$)
\begin{equation}
	\lambda^* := 1 + \frac{2k}{\gamma}(1-\sqrt{1+ \frac{\gamma}{k}}) = 1 - 2 \chi \;.
	\label{eq::lambda_max}
\end{equation}
Therefore, 
\begin{equation}
	\sigma^* := \rho^* - \lambda^* = \chi \;.
	\label{eq::sigma_max}
\end{equation}
Figure~\ref{fig::MF} shows the mean-field solutions for the densities $\lambda$ and $\sigma$ depending on $\rho$, together with the Monte Carlo simulations. The overall agreement between the mean-field approximation and the simulations is very good. Only when $k/\gamma$ is very small the discrepancy between the mean-field and the simulations becomes large (Fig.~\ref{fig::MF}a), as already observed in~\cite{Klumpp:2008cq}. Thus, in the case $k/\gamma \ll 1$, correlations are no longer negligible and the mean-field overestimates the current $J$ (or, equivalently, the amount of particles in the inactive state $n_i=1$). 
However, analytical calculations and simulations show the same qualitative behavior and therefore the mean-field approximation is sufficient to capture the main features of the underlying system. \\
One can see from Eq.~(\ref{eq::rho_max}) and Fig.~\ref{fig::rho_max} that as $k/\gamma$ increases, $\rho^*$ approaches the value $0.5$ and the current profile becomes symmetric with respect to the density $\rho$. This limiting case corresponds to neglecting the internal state of the particles, i.e., considering that the transition $n_i=1 \rightarrow n_i=2$ occurs instantaneously. Note that in this case we recover the results of an ASEP model with a single hopping rate  \nolinebreak $\gamma$.

With increasing $\rho$ there exist different regimes characterized by different amounts of particles in state $1$ or $2$. If $k < \gamma$, then the curves of $\lambda$ and $\sigma$ cross at $\rho_d := 1- k/\gamma$, defining two distinct regimes: for $\rho < \rho_d$, there is  a regime in which  $\lambda >\sigma$, i.e., the density of sites with $n_i=1$ is larger than the density of sites with $n_i=2$. In other words, the mRNA is mainly populated by empty (vacant A site) ribosomes.
For  $\rho > \rho_d$ we have the opposite situation with $\lambda <\sigma$, i.e. the ribosomes have the tRNA in their A site and are waiting to hop. 
These different regimes exist only when $k<\gamma$, otherwise $\lambda <\sigma$ always.
Notice that in general $\rho^* \neq \rho_d$ and therefore, this transition is different from the queueing transition. 
These basic observations, though being a simple study of the ratio between densities, might reveal an interesting biological mechanism (see Section~\ref{sec::conclusions}).

\subsection{Open-boundary conditions \label{ssec::open}}
In this section we focus on the open-boundary conditions and present the corresponding results. We first discuss the outcomes of the model using an iterative map obtained from the mean-field Eqs.~(\ref{eq::densities}). Then we recover the same results using an extremal principle.

A new particle enters the unidimensional lattice with rate $\alpha$ representing the translation inititation. After the injection of a new particle, the first site is set to have $n_1=1$.
As usual, the presence/absence of a particle and its state are represented by the occupation number $n_i$ and in the bulk the dynamics follows the above rules~(\ref{eq::rules}). 
Finally, particles abandon the end of the lattice (when $n_L=2$) with probability per unit time $\beta$ (translation termination). With these prescriptions it is clear that Eqs. (\ref{eq::densities}) hold in the bulk, but have to be modified at the left boundary (injection)
\begin{subequations}
		\begin{align}
			\cfrac{d  \lambda_1}{d t} & =  \alpha(1-  \lambda_1 -  \sigma_1) - k \lambda_1 \\
			\vspace{0.5ex}
			\cfrac{d  \sigma_1}{d t}  & =   k \lambda_1 -  \sigma_{1} (1-\lambda_2-\sigma_2)\gamma \;,
		\end{align}
		\label{eq::sx}
\end{subequations}
and at the right boundary (depletion)
\begin{subequations}
		\begin{align}
			\cfrac{d  \lambda_L}{d t} & =  \sigma_{L-1}(1-  \lambda_L -  \sigma_L)\gamma - k \lambda_L \\
			\vspace{0.5ex}
			\cfrac{d  \sigma_L}{d t}  & =   k \lambda_L -  \beta \sigma_{L} \;.
		\end{align}
		\label{eq::dx}
\end{subequations}
Equations~(\ref{eq::densities}) together with the steady-state condition lead to the following recursive map for the densities $\sigma_i$
\begin{equation}
	\label{eq::recursion}	
	\sigma_{i+1} = 1- J\left(\cfrac{1}{k} + \cfrac{1}{\gamma \sigma_{i}}\right)\;.
\end{equation}
The fixed points of this map are as follows
\begin{equation*}
	\sigma_{\pm} = \frac{1}{2} \left[ \left(1-\frac{J}{k}\right)  \pm  \sqrt{\left(1- \frac{J}{k}\right)^2 - \frac{4J}{\gamma}} \;\right] \;,
\end{equation*}
one of which is stable ($\sigma_+$) and the other unstable ($\sigma_-$).
In an iterative map like Eq.~(\ref{eq::recursion}), $\sigma_{i+1}$ is said to be the homographic function of $\sigma_i$ and, knowing the value of the starting point $\sigma_1$, it is possible to find the general term $\sigma_i$ of the recursion:
\begin{equation}
\label{eq::genterms}
\sigma_i = \cfrac{- \sigma_- \sigma_+ (\sigma_+^{i-1} - \sigma_-^{i-1}) + \sigma_1 (\sigma_+^{i} - \sigma_-^{i})}{- \sigma_- \sigma_+ (\sigma_+^{i-2} - \sigma_-^{i-2}) + \sigma_1 (\sigma_+^{i-1} - \sigma_-^{i-1})} \;.
\end{equation}
Following the approach presented in~\cite{Derrida:1992by} by Derrida and coworkers, we reconstruct the phase diagram of the system by varying the injection rate $\alpha$ and the depletion rate $\beta$ (which are both considered to be smaller than $\gamma$). 

\begin{figure}[tbp]
	\centering
	 \includegraphics[width=0.42\textwidth]{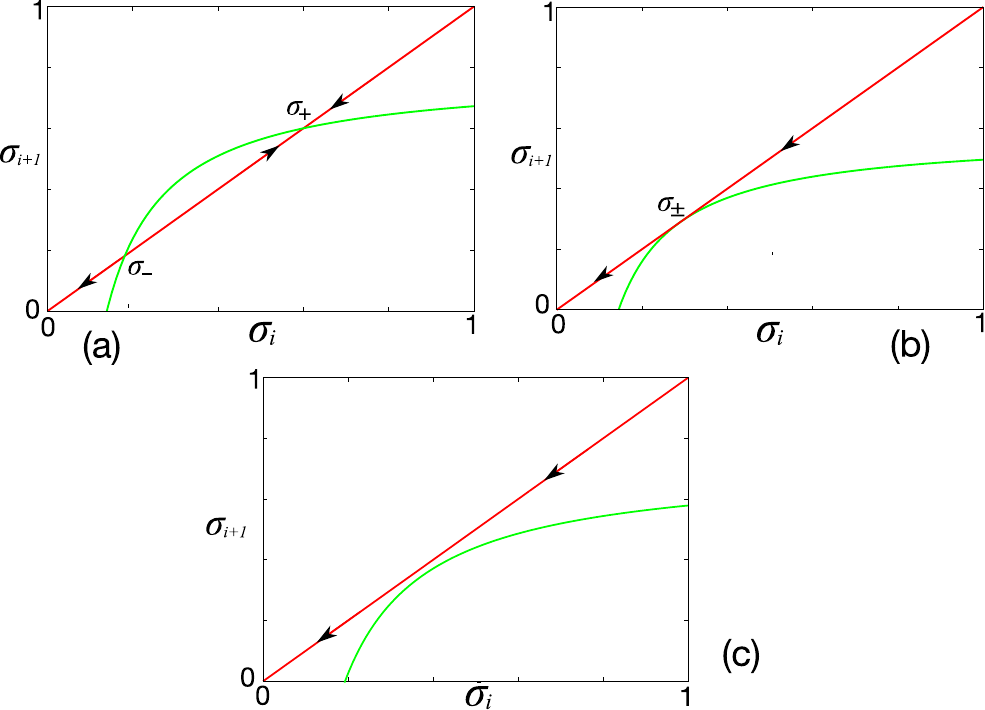}
	 \caption{Graphical representation of the recursive map (\ref{eq::recursion}). (a) When $J<k (1 - 2 \chi)$ there are two different fixed points $\sigma_-$, $\sigma_+$ that collapse when $J=k (1 - 2 \chi)$ (b). Panel (c) shows the map for the finite size case.    \label{fig::rec}}
   \end{figure} 
\begin{figure}[bp]
	\centering
	 \includegraphics[width=0.5\textwidth]{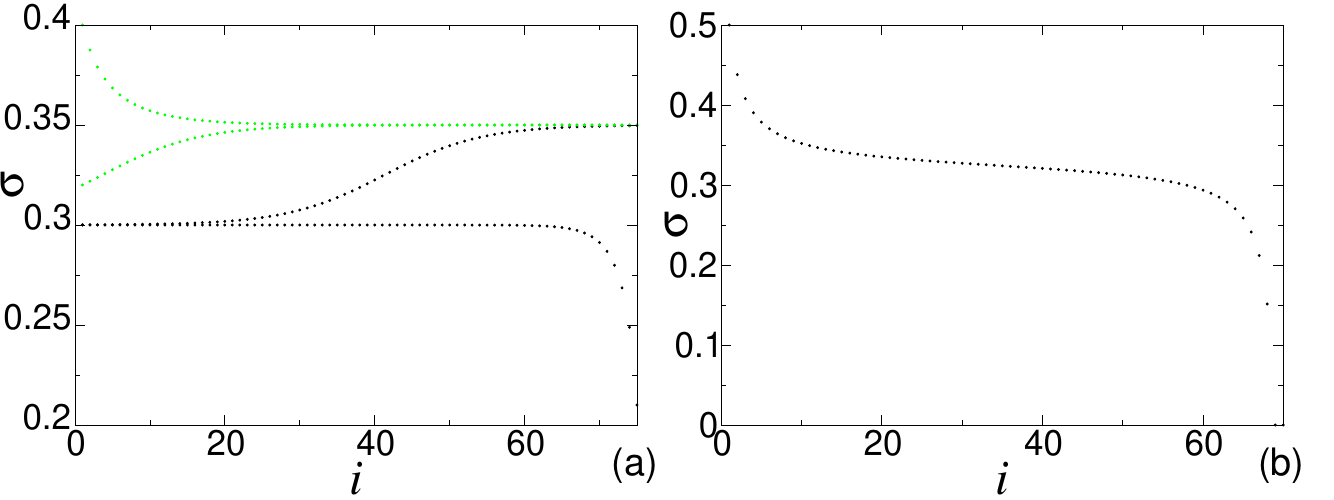}
	 \caption{(Color online). Profiles of the density $\sigma_i$ calculated from Eq.~(\ref{eq::recursion}). (a) When the fixed points exist, the density can either start close to $\sigma_-$ and then go away from this value at the end of the chain (black -lower- dots) or reach the value of the stable fixed point $\sigma_+$ after few iterations [green (light gray) dots]. (b) Typical density profile for the MC region with high density close to the left boundary and low density close to the terminating site.\label{fig::7}}
   \end{figure} 
Using Eq.~(\ref{eq::genterms}) we can calculate $\sigma_L$ as a function of $\sigma_1$ and $J$:
\begin{equation}
	\label{eq::sigmaL}
	\sigma_L = \sigma_L(\sigma_1, J) \;.
\end{equation}
Equation~(\ref{eq::sigmaL}), together with Eqs.~(\ref{eq::sx}a) and~(\ref{eq::dx}b) in the steady-state condition, determines the values of $\sigma_1$, $\sigma_L$ and $J$ as a function of $\alpha$ and $\beta$. 
Moreover, note that in the mean-field approximation the densities $\lambda_i$ can be readily calculated from $\lambda_i = J/k$.
The system shows three different regimes that can be characterised by reasoning on the graphical representation of Eq.~(\ref{eq::recursion}) (see Fig.~\ref{fig::rec}).

\emph{Low Density phase ($\sigma_1 \simeq \sigma_- $,  $\sigma_L < \sigma_+$).} If we start to iterate the map (\ref{eq::recursion}) close to the unstable fixed point $\sigma_-$, at the beginning of the lattice the values of $\sigma_i$ remain close to this value and then move away (black dots in Fig.\ref{fig::7}a). This is the so-called Low Density phase (LD). The recursion (\ref{eq::recursion}) and Eqs.~(\ref{eq::sx}) and~(\ref{eq::dx}) provide the solutions
\begin{align*}
	\sigma_1 & = \frac{\alpha}{\gamma} & \sigma_L = \frac{\alpha k (\gamma - \alpha)}{\beta \gamma (k + \alpha)} \;,
\end{align*}
\begin{equation*}
	J =  \frac{\alpha k (\gamma - \alpha)}{\gamma (k + \alpha)} \;.
\end{equation*}
These equations are valid as long as
\begin{align}
	\label{eq::alpha_c}
	\alpha & \leqslant \alpha_c := \gamma \chi \;, & \beta > \alpha\;,
\end{align}
since otherwise Eqs.~(\ref{eq::sx}),~(\ref{eq::dx}) and~(\ref{eq::recursion}) are not consistent with the conditions $\sigma_1 = \sigma_-$ and $\sigma_L < \sigma_+$. The critical value $\alpha_c$ determines the boundary of the LD regime.

\emph{High Density phase ($\sigma_1 > \sigma_- $,  $\sigma_L \simeq \sigma_+$).} Similarly, the High Density phase (HD) is reached starting from a value $\sigma_1 > \sigma_- $. Iterating the map, we reach a value $\sigma_i$ arbitrarily close to the stable point $\sigma_+$ [green (light gray) dots in Fig.\ref{fig::7}a]. The initial point $\sigma_1$ lies therefore in the domain of attraction of $\sigma_+$. Following the previous procedure, one obtains the solutions 
\begin{align*}
	\sigma_1 & = \frac{\alpha k \gamma - \beta k \gamma + \alpha \beta^2 +k \beta^2}{\alpha \gamma (\beta + k)} & \sigma_L = \frac{ k (\gamma - \beta)}{\gamma (k + \beta)} \;,
\end{align*}
\begin{equation*}
	J  = \frac{ \beta k (\gamma - \beta)}{\gamma (k + \beta)} \;.
\end{equation*}
These solutions exist when 
\begin{align}
	\label{eq::beta_c}
	\beta & \leqslant \beta_c := \gamma \chi \;, & \beta < \alpha \;.
\end{align}
It is worth noting that the critical points $\alpha_c$ and $\beta_c$ delimiting the LD and the HD phases are functions of $k$ and $\gamma$.

\emph{Maximal Current phase ($\sigma_1 \geqslant 2^{-1}(1-J/k) $,  $\sigma_L \leqslant 2^{-1}(1-J/k))$.} This regime is reached when the two fixed points collapse and the lattice carries the maximal current allowed. This phase occurs when 
\begin{align}
	\alpha & \geqslant \alpha_c  \;, & \beta & \geqslant \beta_c \;,
\end{align}
and we have the solutions
\begin{align*}
	\sigma_1 & =1- \frac{J}{k} -\frac{J}{\alpha} & \sigma_L = \frac{J}{\beta} \;,
\end{align*}
\begin{equation*}
	J= k (1 - 2 \chi)\;.
\end{equation*}
We expect these results to hold in the limit $L\rightarrow \infty$ and the finite-size effects to be similar to the ones of standard ASEP~\cite{Derrida:1992by}. Thus, in a finite-size system, the recursion (\ref{eq::recursion}) would not have any real fixed points and the graphical representation of Fig.~\ref{fig::rec}b would have to be modified into Fig.~\ref{fig::rec}c. The role of the limited length $L$ needs further investigation, but this analysis goes beyond the scope of this paper.\\

Until now we have used Eq.~(\ref{eq::recursion}) as the starting point to characterise the different regimes of the process. The \emph{Maximal Current Principle} (MCP) is another viable approach which leads to the same results. It was first presented by Krug~\cite{Krug:1991tg} and then later extended~\cite{Popkov:1999hc, Hager:2001hw}. According to this principle, the boundaries are substituted by reservoirs of particles and the dynamics between the reservoirs and the lattice is assumed to be the same as in the bulk. The MCP states that the current $J$ of an open-boundary lattice in the MC regime is given by
\begin{equation}
	\label{eq::MCP}
	J= \max_{\rho \in \left[ \rho_{L+1}, \rho_0 \right]} J(\rho),
\end{equation}
where $\rho_0$ and $\rho_{L+1}$ are respectively the densities of the reservoirs of particles at the left and the right boundaries. 
 $J(\rho)$ is the expression of the current as a function of the density $\rho$ that in the bulk we can consider to be given by Eq.~(\ref{eq::Jring}).
The densities $\rho_0$ and $\rho_{L+1}$ are chosen to realize the injection and depletion parameters $\alpha$ and $\beta$. 
Equation~(\ref{eq::MCP}) is valid for systems in which the current profile has only one maximum, and has to be modified if $J(\rho)$ presents minima~\cite{Popkov:1999hc, Hager:2001hw}. There are no general prescriptions for choosing the correct densities $\rho_0$ and $\rho_{L+1}$ of the reservoirs~\cite{Shaw:2003xw}. Here we propose a way to fix $\rho_0$ and $\rho_{L+1}$ and relate them to the injection and depletion rate; with these values we recover the results obtained above.

If we imagine having a reservoir of particles or an extra site at $i=0$ with density of particles $\rho_0= \lambda_0 + \sigma_0$, the parameter $\alpha$ can be written as $\alpha = \gamma P(n_0=2)$, where $P(n_0=2)$ is the probability of having the occupation number of the site $i=0$ equal to $2$, i.e., having a particle ready to hop from the reservoir to the lattice. Since $P(n_0=2) = \sigma_0$ one may write~\footnote{It is possible to carry on the procedure with $\rho_0$ and $\rho^*$ instead of $\sigma_0$ and $\sigma^*$ but we find this way straightforward.}:
\begin{equation*}
	\alpha = \sigma_0 \gamma \;.
\end{equation*}
On the right boundary we can assume that the density of particles at the extra-site $L+1$ is related to the depletion rate $\beta$ as follows
\begin{equation*}
	\beta = (1-\rho_{L+1}) \gamma \;.
\end{equation*}
Now, bearing in mind that $\chi = \sigma^*$ and using Eq.~(\ref{eq::rho_max}), the maximal principle yields the location of the critical points by equating $\sigma_0$ with $\sigma^*=\rho^*-\lambda^*$ and $\rho_{L+1}$ with $\rho^*$. 
The transitions occur at the same values $\alpha_c$ and $\beta_c$ obtained before in Eqs.~(\ref{eq::alpha_c}) and~(\ref{eq::beta_c}). The current and the bulk densities are then given by the following equations: 
\begin{equation*}
	J = 
	\begin{cases}
		\vspace{1.5ex}
		 \cfrac{\alpha k (\gamma - \alpha)}{\gamma (k + \alpha)} & \text{for  \( \alpha<\beta< \gamma\chi \) \quad (LD)}\\
		 \vspace{1.5ex}
		  \cfrac{ \beta k (\gamma - \beta)}{\gamma (k + \beta)}  & \text{for \( \beta<\alpha< \gamma\chi \) \quad (HD)}  \\
		  k (1 - 2 \chi ) & \text{for \( \alpha, \beta \geqslant \gamma\chi \) \qquad(MC)}
	\end{cases}
\end{equation*}

\begin{equation*}
	\rho = 
	\begin{cases}
		\vspace{1.5ex}
		 \rho_0 & \text{for \( \alpha<\beta< \gamma\chi \)  \quad (LD)}\\
		 \vspace{1.5ex}
		  \rho_{L+1} & \text{for \( \beta<\alpha< \gamma\chi \)  \quad (HD)}  \\
		  1-\chi   & \text{for \( \alpha, \beta \geqslant \gamma\chi \)  \qquad (MC)} \;.
	\end{cases}
\end{equation*}

\begin{figure}[h!]
	\centering
	 \includegraphics[width=0.29\textwidth]{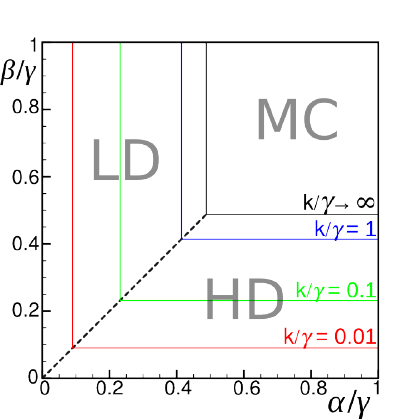}
	 \caption{(Color online). Phase diagram of the ASEP with two-state particles. Different colored lines correspond to different critical values obtained with changing $k/\gamma$. The MC region is larger for low values of $k/\gamma$ and approaches the MC region of the standard ASEP in the limit $k/\gamma \rightarrow \infty$. The dashed line separating the LD and HD denotes the first order transition between these two regions.\label{fig::8}}
\end{figure}

Now we are finally able to construct the rich phase diagram of the system (Fig.~\ref{fig::8}). The model shows the same variety of phase transitions of ``standard'' (particles without internal states) ASEPs, but the borders between the different phases crucially depend on both $k$ and $\gamma$. There are three different regimes (LD, HD, MC) and the transitions towards the MC phase are smooth, i.e., there is a discontinuity in the second derivative of the current profile. On the other hand, the transition between LD and HD is an abrupt transition. The critical points $\alpha_c$ and $\beta_c$ have the same dependency on the parameters $k$ and $\gamma$. The transition line between LD and HD is a straight line in the $\alpha-\beta$ plane and is given by the condition $\alpha=\beta$. Other works taking into account the biochemical cycle of ribosomes~\cite{Garai:2009ta} find a more complicated relation between $\alpha_c$ and $\beta_c$, apparently due to another choice of the densities in the reservoirs. 
However, here we obtain the same results with both the MCP and the mean-field analysis. 
 
For a fixed value of the translocation rate $\gamma$, the boundaries between the MC and the LD and HD regimes are shifted with varying $k$, and the MC phase becomes larger than the MC region of standard ASEPs. 
Importantly, in the limiting case $k \rightarrow \infty$ the critical points approach the values obtained for an ASEP. Since this limit corresponds to consider transitions $n_i=1 \rightarrow n_i=2$ occurring instantaneously by neglecting the internal state of the particles, we therefore confirm that the results obtained with our model are consistent with previous findings. 

\begin{figure}[bthp]
	\centering
	 \includegraphics[width=0.4\textwidth]{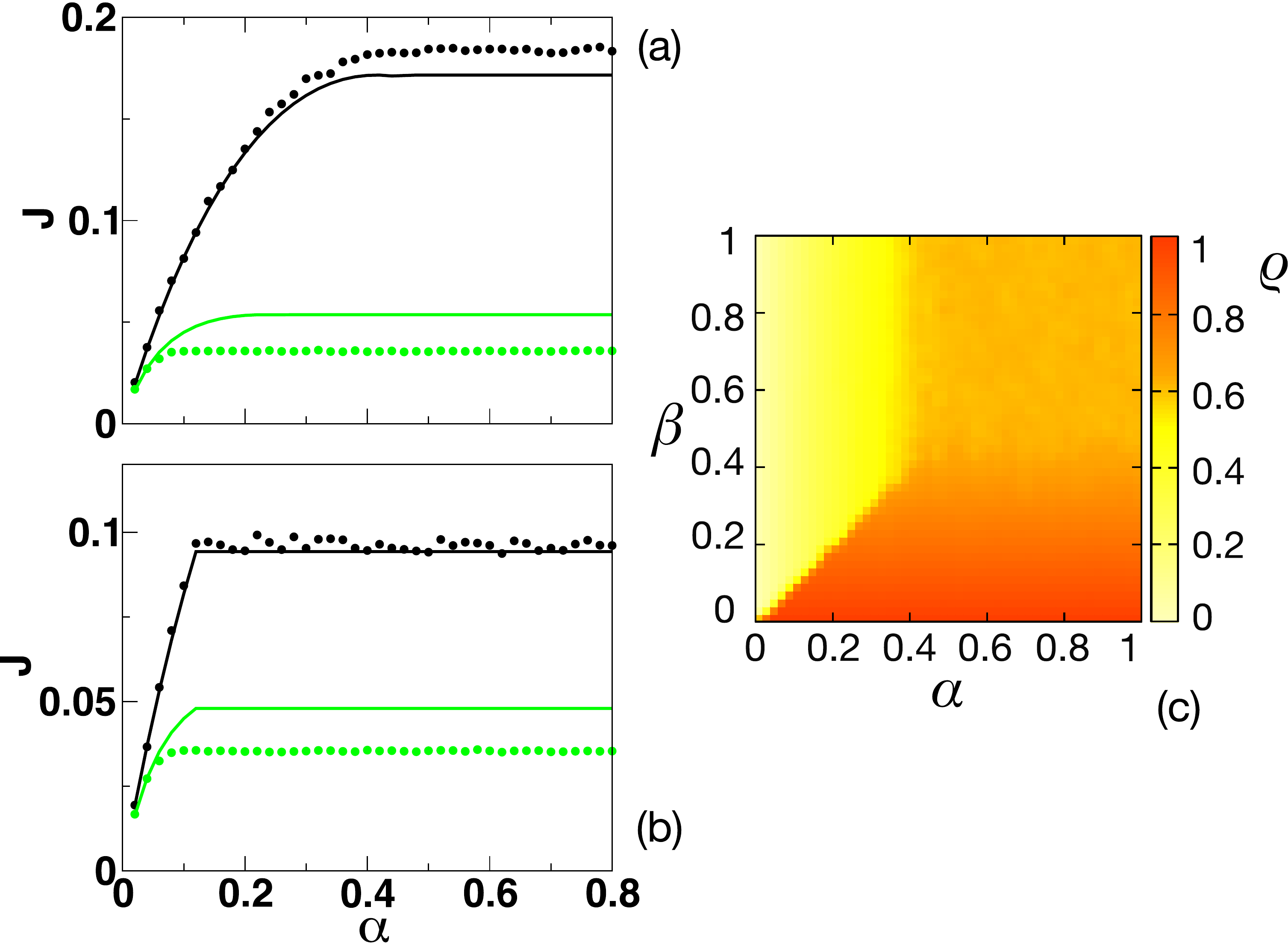}
	 \caption{(Color online). Monte Carlo simulations. Panels (a) and (b) show the current J as a function of $\alpha$ in (a) systems passing from LD to MC ($\beta=1$) and (b) from LD to HD phase ($\beta=0.12$). Full lines represent the theoretical predictions and circles are simulation points ($k=1$ in black and $k=0.1$ in green - light gray). The value of $\gamma$ is fixed to $1$. Panel (c) shows the numerical phase diagram (different colors represent different densities $\rho$) for a system with $k=\gamma=1$. Drawing the phase diagram for other values of the parameters, we reproduce the features illustrated in Fig.~\ref{fig::8}.\label{fig::9}}
\end{figure}

Figure~\ref{fig::9} shows the outcomes of numerical simulations. As in the close-boundary case, the analytical results present deviations from the Monte Carlo simulations for low values of $k/\gamma$. Despite that, a study of the numerical phase diagram (Fig.~\ref{fig::9}c) shows the same phenomenology of the analytical treatment of Fig.~\ref{fig::8}. Only the location of the critical points is inaccurate in the mean-field theory. For instance, notice that for some values of the parameters, in Fig.~\ref{fig::9}b the mean-field predicts a LD-HD transition instead of the smooth LD-MC transition numerically found, i.e., the MC phase is reached for lower values of $\alpha$ and $\beta$ (green circles). In other words, the numerically observed MC region is even larger than the one predicted by the mean-field approximation. 

\section{Discussion and Conclusions}
\label{sec::conclusions}

In this work we have proposed a model for mRNA translation based on an exclusion process. We have extended previous models based on the ASEP (Asymmetric Simple Exclusion Process) by including the internal stepping cycle of the ribosomes, which corresponds to allowing the particles to have multiple internal states.
The same model has been previously introduced in~\cite{Klumpp:2008cq} to study the traffic of molecular motors.
We have condensed the whole biochemical cycle of the ribosome into two main steps: (i) finding the correct tRNA, which occurs with rate $k$, and (ii) translocation rate of the ribosome to the next codon, which happens with rate \nolinebreak$\gamma$. 

This extension is crucial in describing the underlying biological process, since previous ASEP-based models neglect that ribosomes can find and keep a correct tRNA during the waiting time due to the occupation of the next codon. 

The main result of this work is that the transitions among the different dynamical regimes of the system occur at different critical points than the ones predicted by former ASEP models. These critical points depend on both $k$ and $\gamma$. For example, when $k/\gamma$ is small, the MC phase is substantially enlarged compared to a standard ASEP with hopping rate $\gamma$ and which ignores the internal degree of freedom ($k \rightarrow \infty$)~\footnote{However, one might prefer to compare the model discussed here with a standard ASEP having an hopping rate $p= \gamma k / (\gamma +k)$, i.e. the inverse of the average waiting time on a site for the two state model. In this case, by decreasing $k$ one obtains a MC region that is \emph{bigger} than the one obtained with the two state model studied in this work. We thank Paul Higgs for this useful comment.}. Crucially, this is the biologically relevant regime, as shown by estimates of the parameters based on experimental data. Thus, this model describes the biological system much more accurately, and its predictions can be readily validated by experimental measurements. 
 
The analysis of the system with periodic boundary conditions introduces the model and the formalism. This situation has been studied in~\cite{Klumpp:2008cq} where the authors propose a mean-field approach to analyse the case with periodic boundary conditions. Here we present a more complete approach which recovers the previous results and, in addition, makes possible the treatment of the model in the open boundary case. When the ratio between the transition rates $k$ and $\gamma$ is high, the current profile becomes symmetric and the value of $\rho^*$ moves toward the expected value of a standard exclusion process. The deviation between the mean-field and our simulations for low values of the transition rate $k$ has yet to be understood. Furthermore, our model with periodic boundary conditions allows us to study whether there is a dominance of particles in state $1$ or $2$. In the biological system that we describe (even if in a coarse-grained perspective), a lattice with the majority of sites having $n_i=2$ represents an mRNA in which ribosomes are carrying the cognate tRNAs and are waiting for hopping. This might be unfavourable when a finite number of charged tRNAs is available. Roughly speaking, in these conditions the charged tRNAs are kept by the ribosomes and cannot be used to translate other codons. In this sense the usage of resources is not optimized if $\sigma > \lambda$. This result suggests that under starvation or stress conditions, there might be a transition from the $\sigma > \lambda$ to the $\sigma < \lambda$ regime.

We have shown that this extension of the model has important consequences for the different boundary-induced transitions. Namely, depending on the ratio of $k$ and $\gamma$, the sizes of the low density (LD), high density (HD) and maximal current (MC) phases in the $\alpha-\beta$ parameter space can change substantially, where $\alpha$ and $\beta$ represent respectively the initiation and termination rate of ribosomes. Crucially, the phase diagram coincides with the one obtained with the ASEP if $k/\gamma \gg 1$, whereas if $k/\gamma \rightarrow 0$, the maximal current phase is enlarged to a great extent, and the transitions from the LD and HD to the MC phase occur at much lower values of $\alpha$ and $\beta$ (depending on the value of $k$, simulations show that the critical points are overestimated by the mean-field approach and the MC region is even larger than predicted).

Based on experimental data, the translocation rate $\gamma$ is estimated to be $\gamma = 35$ s$^{-1}$  ~\cite{Savelsbergh:2003qc} (which is naturally assumed to be constant for each codon), and the ratio $k/\gamma$ turns out to be in the range $0.05-3.38$, depending on the codon (these values are estimates based on~\cite{Gilchrist:2006wd}). For most codons ($\simeq 87\%$) the ratio of the rates is smaller than 1.  Therefore, for physiological conditions, our model predicts a much larger MC phase in the parameter space than previous ASEP models which completely neglect the internal state of the ribosomes or, equivalently, assume that $k/\gamma \gg 1$ contrary to the biological conditions. In that unrealistic situation, the ribosomes would find the cognate tRNAs as soon as they translocate to the next codon. 

Thus one could naively think that the translation process is optimized to produce the largest possible number of proteins per unit time. In other words, the current $J$ is maximized, and hence the system is in the MC phase. Although this assertion is not justified and there might be cases in which other effects prevent the translation rate to become maximal, such as particular configurations of slow codons downstream of the 5' end of the mRNA, 
competition for common resources and regulation at the level of translation, in this work we find signatures in this direction. In particular, the model predicts that the MC phase occupies a very large region in the parameter space. Furthermore, the density profiles experimentally observed~\cite{Ingolia:2009dq} recall the MC density profile of Fig.~\ref{fig::7}b. Finally, ribosome recycling~\footnote{If the 5' and 3' ends of mRNAs interact with each other, then the polynucleotide chain is deformed and ribosomes which have just finished translating the protein can be quickly re-used (\emph{recycled}). Thus, the local concentration of free ribosomes at the 5' cap increases proportionate to the current $J$.} might drive the system to lie in the region with the highest current. In fact, the injection parameter $\alpha$ can be decomposed into two components: one constant coefficient $\alpha_{o}$ being the affinity of freely diffusing ribosomes to bind to an open mRNA, and an increasing function of $J$  which represents the probability per unit time that a ribosome leaving the end of the mRNA will be recycled. It is clear that $\alpha$ might grow until the current balances the maximal current $J^*$ and as a result, $\alpha$ is larger or equal to the critical value $\alpha_c$.
The ribosome recycling and its potential impact on translation regulation has been thoroughly investigated in~\cite{Chou:2003qd}.

Further studies on the model proposed in this work will address lattices with inhomogeneities (slow codons) and the influence of the size of the particles (ribosomes are known to cover around 9 codons). On the biological side, we are planning to implement simulations of real mRNA sequences from the \emph{S. cerevisiae} genome and validate the model with experiments at different levels.

\begin{acknowledgments}
\end{acknowledgments}
The authors thank R.~J. Allen and M.~R. Evans for useful discussions during the development of the model. One author (L.~C.) would also like to thank A. Parmeggiani for constructive correspondence and comments of previous versions of the manuscript. This work has been supported by SULSA and BBSRC [BB/F00513X/1 and BB/G010722].


\end{document}